\documentclass{ws-procs9x6}
\usepackage{graphics,subfigure}

\begin{document}

\title{The SuperNEMO double beta decay experiment}

\author{I. Nasteva$^*$ on behalf of the SuperNEMO Collaboration}

\address{School of Physics and Astronomy, University of Manchester,\\
Oxford Road, Manchester, M13 9PL, UK\\
$^*$E-mail: Irina.Nasteva@manchester.ac.uk}

\begin{abstract}
The SuperNEMO project studies the feasibility of employing a technique of tracking plus calorimetry
to search for neutrinoless double beta decay in 100~kg of enriched isotopes. It aims to reach an 
effective neutrino mass sensitivity of 50~meV\@. The current status of the SuperNEMO R\&D programme 
is described, focusing on the main areas of improvement.
\end{abstract}


\bodymatter

\section{Introduction}\label{aba:sec1}
Neutrinoless double beta decay is the most sensitive process for probing lepton number violation. 
Its discovery would prove that neutrinos are Majorana particles, and would give access to their 
mass scale.

NEMO~3\cite{nemo3tdr,nemoresult1,nemoresult2,nemoresult3} is a neutrinoless double beta decay experiment currently running in 
the Modane Underground Laboratory. 
Based on its experience and 
technological expertise, the SuperNEMO project investigates the design of the next generation 
double beta decay experiment. 

\section{Goals of the SuperNEMO project}
SuperNEMO aims to extend and improve the experimental techniques 
used by the current NEMO~3 experiment to search for neutrinoless double beta decay. 
It will extrapolate the successful NEMO~3 technology over one order of magnitude by studying
about 100~kg of double beta decay isotopes, probably $^{82}$Se or $^{150}$Nd. 
The experiment will benefit from a relatively short R\&D programme, 
and its first modules can start taking data in 2010. 
By 2015, SuperNEMO will reach a sensitivity on the half-life of neutrinoless double beta decay 
of $T_{1/2}>2\times 10^{26}$~years, corresponding to an effective neutrino mass reach of $\langle m_{\nu}\rangle <50-90$~meV\@.

\section{SuperNEMO detector design}

\begin{figure}[tb]
\subfigure{\epsfxsize=6.5cm \epsffile{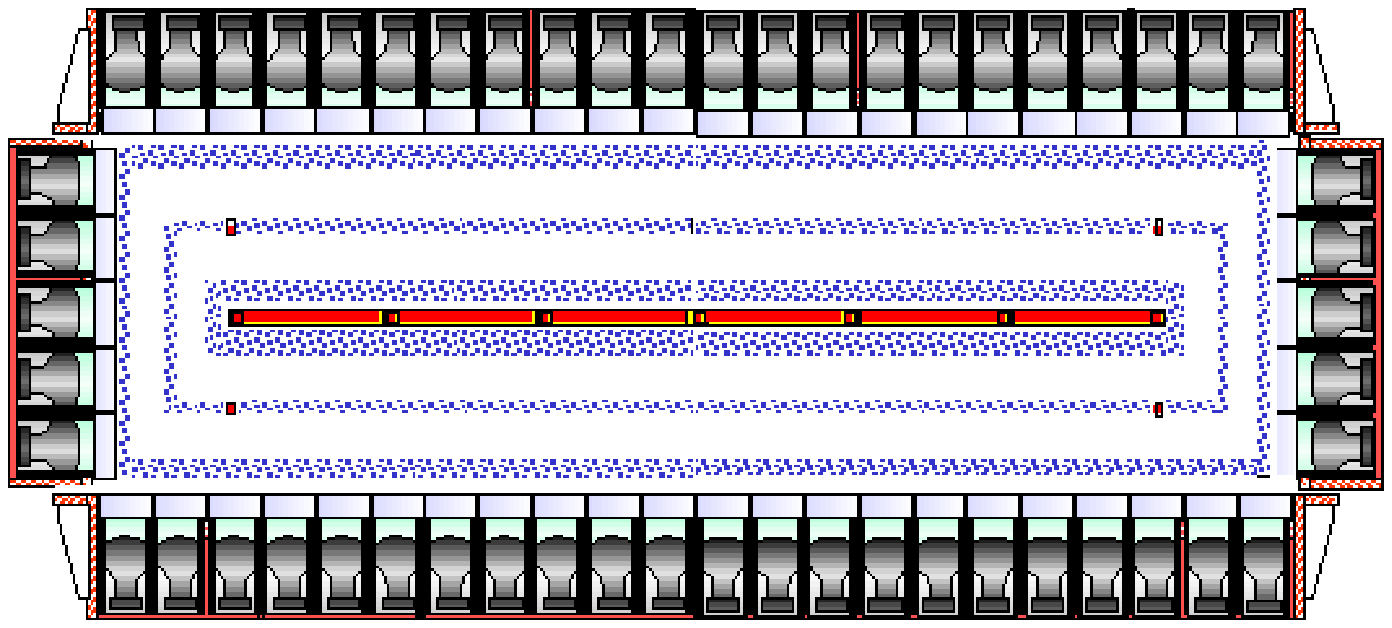}}
\hspace{1cm}
\subfigure{\epsfxsize=3.7cm \epsffile{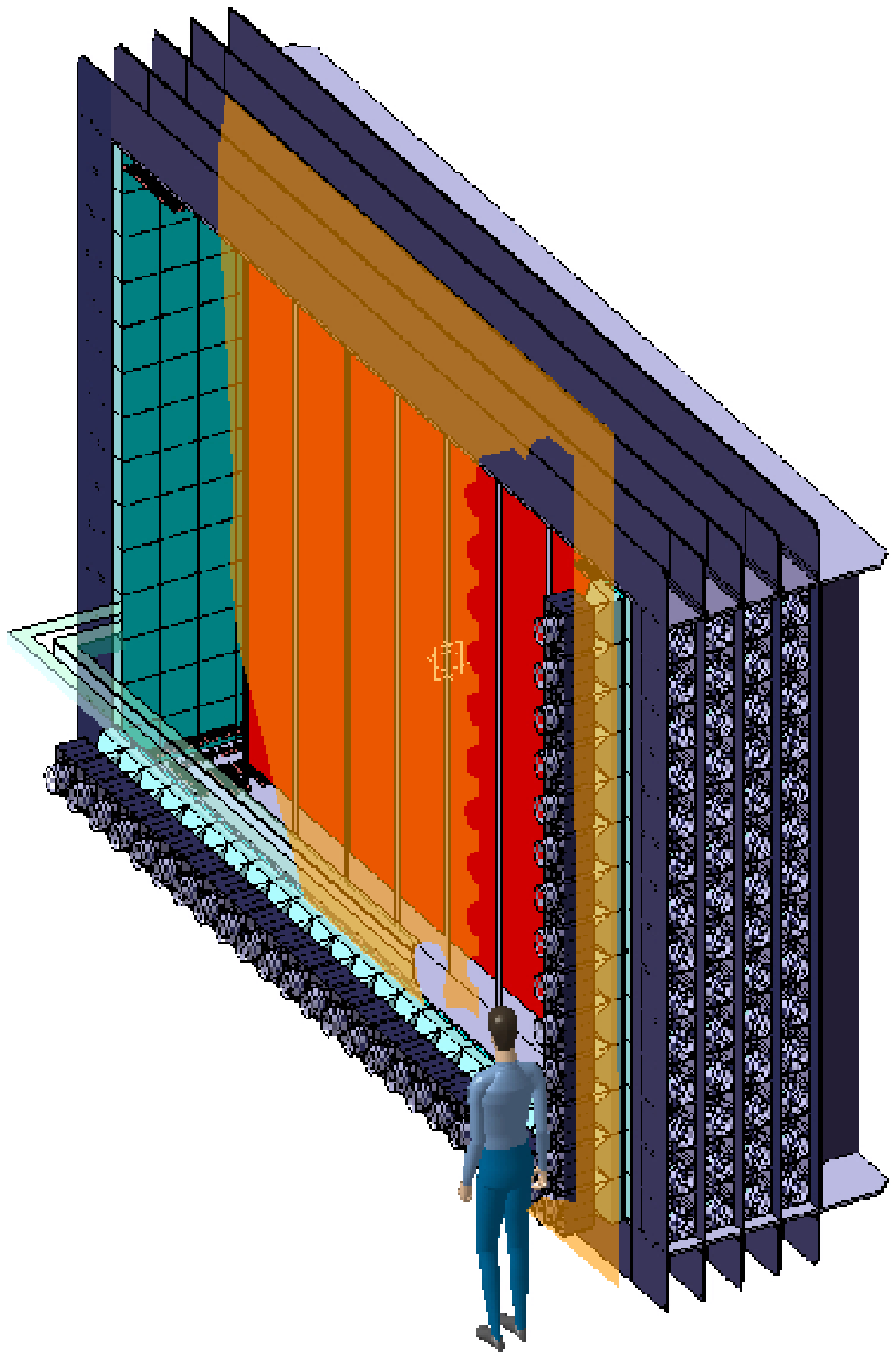}}\hspace{1cm}\\
\caption{Preliminary design of the SuperNEMO detector: view from above (left) and three-dimensional cut-away view (right).}
\label{snemo}
\end{figure}

To fulfil these physics goals, SuperNEMO will build upon the NEMO~3 technology choice of 
combining calorimetry and tracking. 
This gives the ability to measure individual electron tracks, vertices, energies and time 
of flight, and to reconstruct fully the kinematics and topology of an event. 
Particle identification of gamma and alpha particles, as well as distinguishing electrons from
positrons with the help of a magnetic field, form the basis of background rejection. 
An important feature of NEMO~3 which is kept in SuperNEMO is the fact that the double beta decay 
source is separate from the detector, allowing several different isotopes to be studied. 

SuperNEMO will consist of about twenty identical modules, each housing around 5~kg of isotope. 
A preliminary design of a SuperNEMO detector module is shown in Fig.~\ref{snemo}.
The source is a thin ($\sim 40$~mg/cm) foil inside the detector. 
It is surrounded by a gas tracking chamber followed by calorimeter walls. 
The tracking volume contains more than 2000 wire drift chambers operated in Geiger mode, 
which are arranged in nine layers parallel to the foil. 
The calorimeter is divided into $\sim$1000 blocks which cover most of the detector outer 
area and are read out by photo multiplier tubes (PMT).

\section{Status of the SuperNEMO R\&D programme}
The SuperNEMO collaboration was formed in 2005 with the goal of carrying out a three-year 
design study programme (2006--2009) and producing a Technical Design Report (TDR) as outcome. 
The SuperNEMO R\&D programme is currently underway, supported by funding agencies in France, 
the UK and Spain. 

The expected improvement in performance of SuperNEMO compared to its predecessor NEMO~3 
is shown in Table~\ref{parameters} which compares the parameters of the two experiments. 
The most important design study tasks are described in the sections that follow. 

\begin{table}[tb]
\tbl{Comparison of the main NEMO~3 and SuperNEMO parameters.}
{\begin{tabular}{@{}lcc@{}}\toprule
& NEMO 3 & SuperNEMO  \\ \colrule
isotope & $^{100}$Mo & $^{150}$Nd or $^{82}$Se \\  
mass    & 7 kg & 100--200 kg \\
signal efficiency & $8\%$ & $>30\%$ \\
$^{208}$Tl in foil & $<20 \mu$Bq/kg  & $<2 \mu$Bq/kg \\
$^{214}$Bi in foil & $<300 \mu$Bq/kg  & $<10 \mu$Bq/kg ($^{82}$Se) \\
energy resolution at 3 MeV & $8\%$ (FWHM)  & $4\%$ (FWHM) \\
half-life  & $T_{1/2}^{0\nu}>2\cdot 10^{24}$ years & $T_{1/2}^{0\nu}>2\cdot 10^{26}$ years \\
neutrino mass & $\langle m_{\beta\beta} \rangle < 0.3 - 1.3$ eV & $\langle m_{\beta\beta} \rangle < 50 - 90$ meV \\ \botrule
\label{parameters}
\end{tabular}}
\end{table}

\subsection{Energy resolution}
An excellent energy resolution is vital for discriminating the neutrinoless double beta decay peak 
in the energy sum of the two electrons from the background of two-neutrino beta decay. 
The energy resolution is determined by two factors, the electron energy losses in the foil
and the calorimeter resolution. 

SuperNEMO aims to improve the calorimeter energy resolution to $7\%/\sqrt{E}$ at FWHM 
(compared to the NEMO~3 resolution of $\sim 14\%/\sqrt{E}$). 
To reach this goal, several ongoing studies are investigating the choice of calorimeter parameters 
such as scintillator material (organic plastic, liquid, or non-organic), and the shape, 
size and coating of calorimeter blocks. 
These are combined with dedicated development of PMTs with low radioactivity 
and high quantum efficiency. 
Initial studies obtain excellent resolution ($\sim 6.5\%$ at 1~MeV) for calorimeter samples of 
small size; the focus now is to retain this property in larger blocks.

\subsection{Tracker design}

\begin{figure}[tb]
      \subfigure{\epsfxsize=3.2cm \epsffile{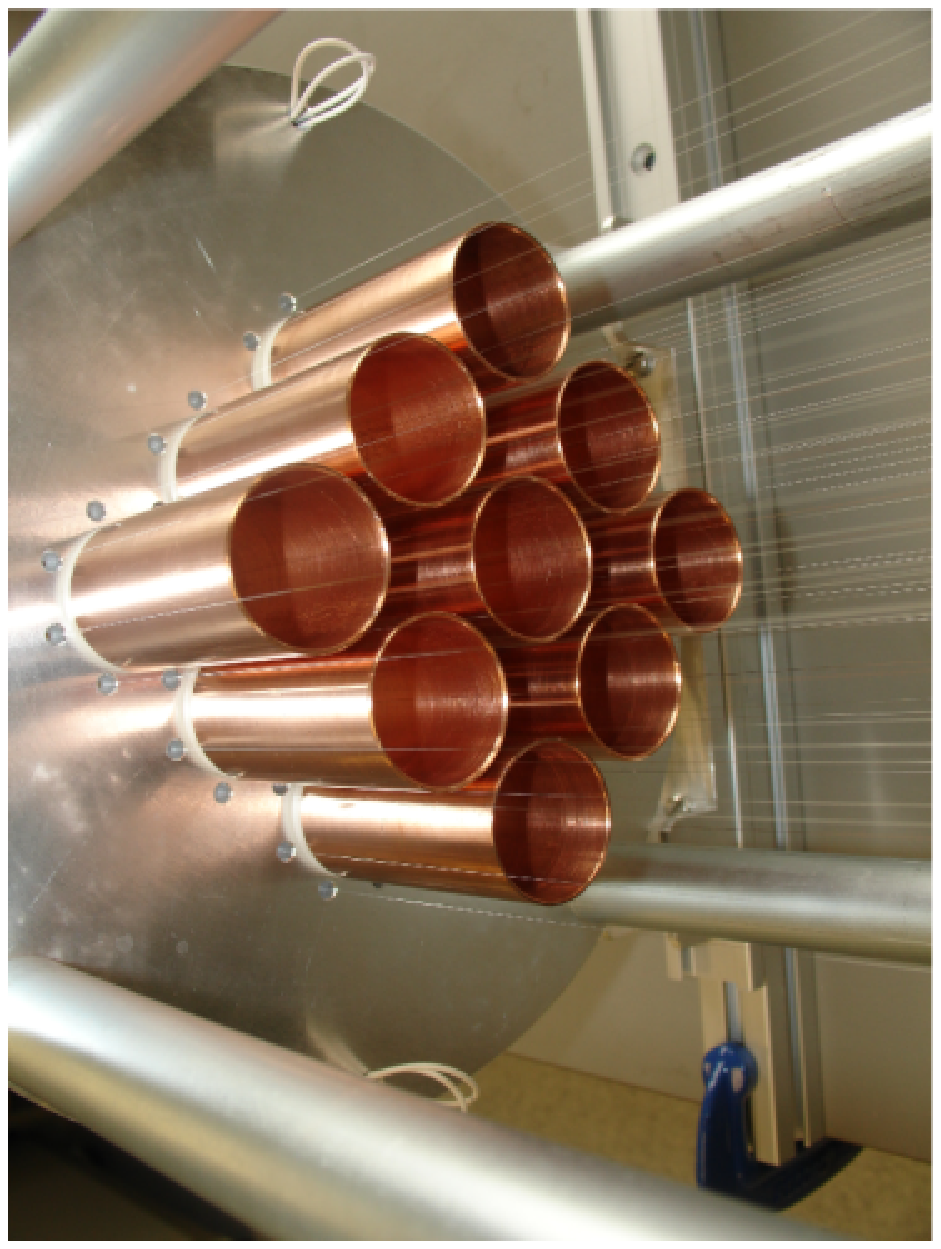}}
\hspace{0.3cm}
\subfigure{\epsfxsize=7.7cm \epsffile{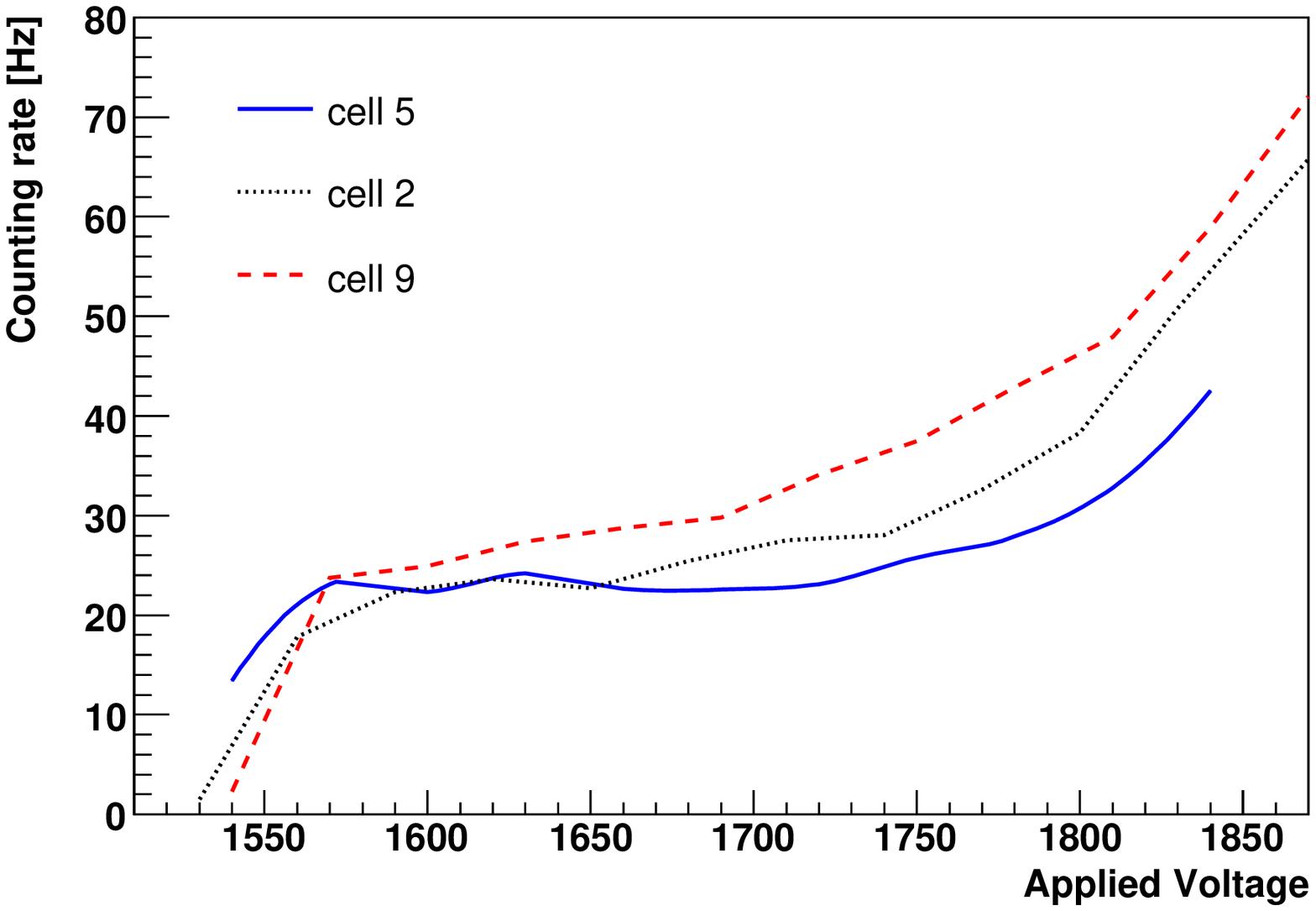}}\hspace{1cm}\\
\caption{The 9-cell tracker prototype. On the left is a photo of the prototype endplate, containing the cathode rings and the terminations of the anode and ground wires. 
 On the right is a plot of the counting rate dependence on anode voltage.}
\label{9cell}
\end{figure}

The SuperNEMO tracker consists of octagonal wire drift cells operated in Geiger mode. 
Each cell is around 4~m long and has a central anode wire 
surrounded by 8--12 ground wires, with cathode pickup rings at both ends. 
Signals can be read out from the anode and/or cathodes to determine the position 
at which the ionising particle crossed the cell.  

The tracking detector design study looks at optimising its parameters to obtain high efficiency 
and resolution in measuring the trajectories of double beta decay electrons, as well as of 
alpha particles for the purpose of background rejection. 
The tracking chamber geometry is being investigated with the help of detector simulations to compare
the different possible layouts. 
In addition, several small prototypes have been built to study the drift chamber cell design 
and size, wire length, diameter and material, and gas mixture. 

The first 9-cell prototype is shown in the photo on the left of Fig.~\ref{9cell}. 
The prototype was successfully operated with three different wire layouts, demonstrating 
a plasma propagation efficiency close to 100\% over a wide range of voltages 
(the Geiger regime operational plateau shown in the flat region on the right plot of Fig.~\ref{9cell}). 
The next step is to construct a 100-cell prototype 
which will test the mechanics and large-scale operation of the drift cell system. 

A SuperNEMO module will contain several thousand drift cells with 8--12 wires each. 
The large total number of wires has lead to the need of automated wiring. 
A dedicated wiring robot is being developed for the mass production of drift cells. 

\subsection{Choice of source isotope}
The choice of isotope for SuperNEMO is aimed at maximising the neutrinoless signal 
over the background of two-neutrino double beta decay and other nuclear decays mimicking the process. 
Therefore the isotope must have a long two-neutrino half-life, and high endpoint energy 
and phase space factor $G_{0\nu}$ ($T_{1/2}^{0\nu} \sim G_{0\nu}^{-1}$). 
The enrichment possibility on a large scale is also a factor in selecting the isotope. 
The main candidate isotopes for SuperNEMO have emerged to be $^{82}$Se and $^{150}$Nd. 
A sample of 4~kg of $^{82}$Se has been enriched and is currently undergoing purification. 
The SuperNEMO collaboration is investigating the possibility of enriching large amounts of $^{150}$Nd 
via the method of atomic vapour laser isotope separation. 

\subsection{Radiopurity of the source}
SuperNEMO will search for a very rare process, therefore it must maintain ultra-low background levels. 
The source foils must be radiopure, and their contamination with naturally radioactive elements must 
be precisely measured. 
The most important source contaminants are $^{208}$Tl and $^{214}$Bi, whose decay energies are close to 
the neutrinoless signal region. 
In order to evaluate their activities, a dedicated BiPo detector is being developed which can measure their 
signature of an electron followed by a delayed alpha particle. 
The first BiPo prototype was installed underground in 2006.

\section{Conclusion}
An extensive R\&D programme is underway to design the next-generation neutrinoless double beta decay 
experiment SuperNEMO. 
It will extrapolate the successful technique of calorimetry plus tracking to 100~kg of source isotope, aiming to 
reach a neutrino mass sensitivity of 50~meV\@. 
Due to its modular approach, SuperNEMO can start operation in stages, with the first module installed as early 
as 2010, and all twenty modules running by 2012--2013.

\bibliographystyle{ws-procs9x6}
\bibliography{supernemo}

\end{document}